\documentclass[aip,amsfonts,amssymb,amsmath,reprint,nofootinbib,twoside,floatfix]{revtex4-1}

\usepackage[english]{babel}
\usepackage[utf8]{inputenc}
\usepackage[T1]{fontenc}
\usepackage{amsthm}
\usepackage{mathtools}
\usepackage{physics}
\usepackage{xcolor}
\usepackage{graphicx}
\usepackage[left=23mm,right=13mm,top=35mm,columnsep=15pt]{geometry}
\usepackage{adjustbox}
\usepackage{placeins}
\usepackage{csquotes}
\usepackage{chemformula}
\usepackage[pdftex,pdftitle={Microwave Stimulated Serpentinization},pdfauthor={Pokharel and Musho}]{hyperref}

\begin{document}

\title{Microwave-Stimulated Serpentinization of Olivine for Geological Hydrogen Production}

\author{Ansan Pokharel}
\author{Terence Musho}
\email[Correspondence email address: ]{tdmusho@mail.wvu.edu}
\affiliation{West Virginia University, Department of Mechanical, Materials, and Aerospace Engineering, Morgantown, WV, USA.}

\date{\today}

\begin{abstract}
Serpentinization of ultramafic rocks is a naturally occurring mineralogical process that can generate molecular hydrogen through the oxidation of ferrous iron during water-rock reaction. Although the resource potential is large, the natural reaction is kinetically limited, and practical hydrogen recovery requires methods that can accelerate conversion without imposing an energy penalty that exceeds the value of the hydrogen produced. This short communication reports a preliminary atmospheric-pressure microwave serpentinization experiment using a water-saturated 2 g crushed olivine sample. Microwave irradiation produced a rapid increase in measured hydrogen concentration compared with conventional hot-plate heating under otherwise similar conditions. The preliminary experiment showed approximately a 12-fold increase in hydrogen concentration and an apparent rate increase from about 2 ppb s$^{-1}$ for conventional heating to about 10 ppb s$^{-1}$ during microwave exposure. These results suggest that electromagnetic stimulation can enhance serpentinization kinetics, likely through rapid volumetric heating, selective coupling to iron-bearing phases, and localized thermal gradients. The result provides an initial experimental basis for evaluating microwave stimulation as a route to accelerated geologic hydrogen production and motivates follow-on measurements using calibrated gas analysis, absorbed-power measurements, dielectric characterization, and elevated-pressure testing.
\end{abstract}


\maketitle

\section{Introduction}

Global energy demand remains strongly coupled to fossil-fuel production, which motivates new low-carbon energy carriers and new methods for producing hydrogen without direct carbon dioxide emissions. Hydrogen is attractive because it can be used in fuel cells, chemical manufacturing, refining, combustion systems, and long-duration energy storage. However, most hydrogen is still produced from fossil feedstocks, most commonly by steam methane reforming or related hydrocarbon conversion processes~\cite{pareek2020insights,international2022global}. Water electrolysis can produce low-emission hydrogen when powered by low-carbon electricity, but cost, efficiency, materials demand, and infrastructure remain major barriers to large-scale deployment~\cite{mazloomi2012electrical,ajanovic2022economics}.

Natural and stimulated geologic hydrogen provide an additional pathway. Serpentinization is a water-rock reaction involving ultramafic minerals such as olivine and pyroxene. During this process, ferrous iron is oxidized, secondary minerals such as serpentine, brucite, and magnetite form, and molecular hydrogen is produced~\cite{berndt1996reduction,schrenk2013serpentinization,lamadrid2017effect}. Estimates of the hydrogen potential associated with ultramafic rocks are substantial, but the rate of hydrogen generation under natural conditions is often too slow for direct energy production without stimulation~\cite{kelemen2011rates,osselin2022orange,worman2016global}. Thus, the central technical challenge is not only the thermodynamic feasibility of hydrogen generation but also the ability to increase reaction rates in a controlled and energy-efficient manner.

Microwave heating offers a possible route for controlled stimulation of serpentinization. Unlike conventional heating, which transfers heat from the outside of a reactor or rock volume inward, microwave energy can be deposited volumetrically in dielectric and lossy mineral phases. This mechanism can create rapid heating, selective heating, and local hot spots in multiphase materials~\cite{agrawal2010latest,peng2015microwave}. These features have been used in microwave-assisted materials processing and microwave-enhanced chemical reactions, including carbothermal reduction of metal oxides and other mineral transformations~\cite{ford1967high,omran2020microwave,huang2012non,hassine1995synthesis}. For serpentinization, the presence of iron-bearing phases and magnetite formation suggests that electromagnetic coupling may evolve during reaction, which could create feedback between mineral transformation and microwave absorption.

\begin{figure}[!]
\centering
\includegraphics[width=0.9\columnwidth]{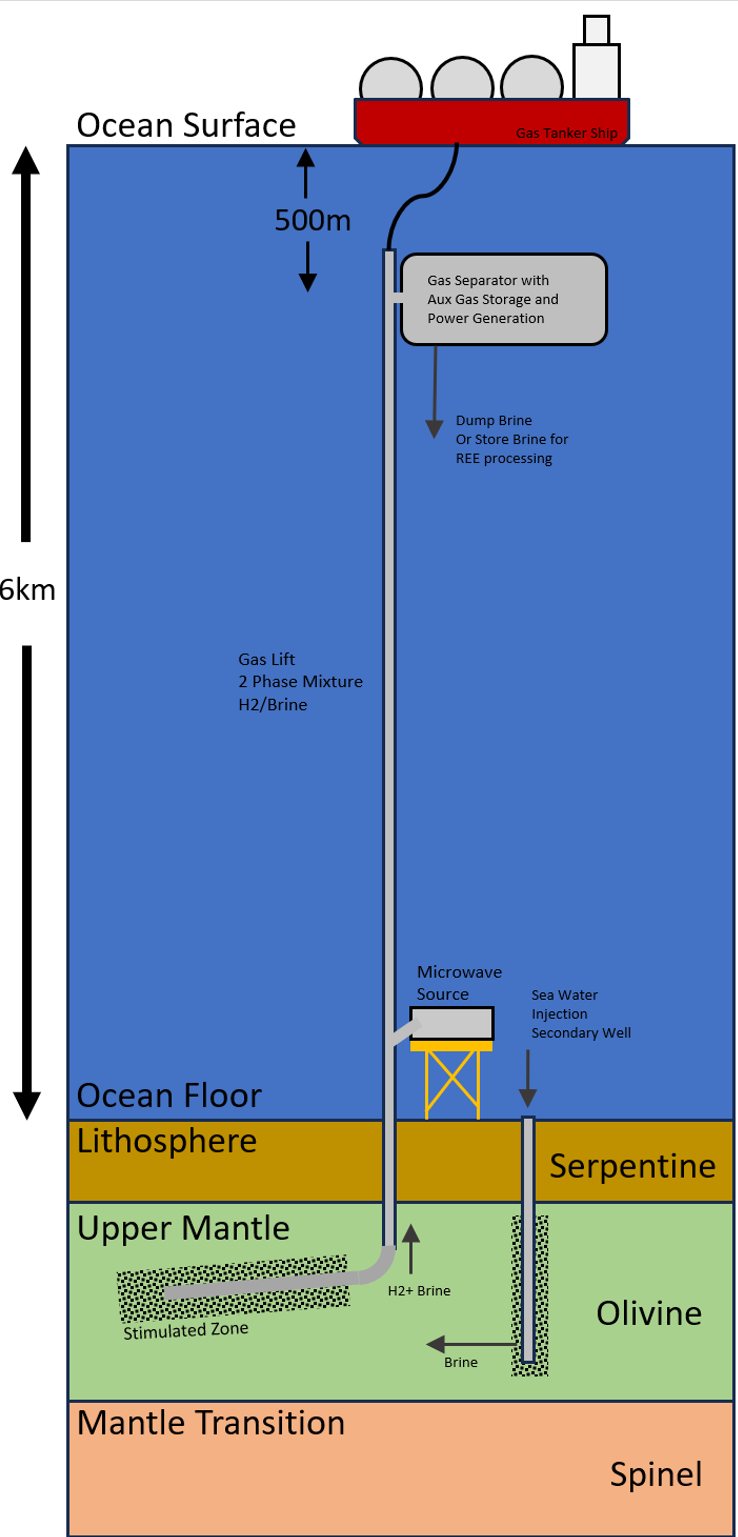}
\caption{Conceptual field configuration for microwave stimulated serpentinization. Radio-frequency energy is delivered through a well-bore geometry to stimulate an olivine-rich formation, while a brine and hydrogen mixture is transported toward the surface.}
\label{fig:concept}
\end{figure}

In this study, we focus on the first experimental step: demonstrating whether microwave exposure can measurably increase hydrogen evolution from a water-saturated olivine sample under controlled laboratory conditions. Figure~\ref{fig:concept} illustrates the broader motivation for the work. At larger scale, radio-frequency energy could be delivered through a well-bore geometry to stimulate an olivine-rich formation, while a secondary well provides water and a gas-lift or related artificial-lift method transports a brine and hydrogen mixture toward the surface. The immediate technical question addressed here is narrower: whether microwave irradiation produces a higher hydrogen response than conventional heating for a small crushed olivine sample at atmospheric pressure.
\section{Serpentinization reaction}\label{sec:serpentinization}

Serpentinization is a general term for the hydration and alteration of ultramafic minerals. For olivine-rich systems, the reaction can be written in simplified form as
\begin{multline}
\mathrm{olivine} + \mathrm{water} \rightarrow
\mathrm{serpentine} + \mathrm{brucite} \\
+ \mathrm{magnetite} + \mathrm{hydrogen}.
\end{multline}
A more specific representation for an Fe-bearing olivine composition is
\begin{widetext}
\begin{multline}
\mathrm{Mg}_{1.82}\mathrm{Fe}_{0.18}\mathrm{SiO}_4 + w\mathrm{H}_2\mathrm{O} \rightarrow
0.5(\mathrm{Mg},\mathrm{Fe}^{II},\mathrm{Fe}^{III})_3(\mathrm{Si},\mathrm{Fe}^{III})_2\mathrm{O}_5(\mathrm{OH})_4 \\
+ x(\mathrm{Mg},\mathrm{Fe})(\mathrm{OH})_2 + y\mathrm{Fe}_3\mathrm{O}_4 + z\mathrm{H}_2 .
\label{eq:serpentinization_general}
\end{multline}
\end{widetext}
Hydrogen generation is associated primarily with oxidation of ferrous iron and the formation of ferric iron-bearing phases. A simplified redox expression is
\begin{equation}
2\mathrm{Fe}^{II}\mathrm{O}+\mathrm{H}_2\mathrm{O}\rightarrow \mathrm{Fe}^{III}_2\mathrm{O}_3+\mathrm{H}_2 .
\label{eq:redox}
\end{equation}

Previous experimental studies show that hydrogen production depends on temperature, pressure, pH, water activity, mineral composition, and reaction time. McCollom \textit{et al.} studied hydrogen generation and iron partitioning during experimental serpentinization of olivine-pyroxene mixtures and observed long induction periods followed by increased hydrogen production under alkaline conditions~\cite{mccollom2020hydrogen}. Klein \textit{et al.} showed that olivine composition, especially iron content, strongly affects hydrogen generation potential~\cite{klein2013compositional}. McCollom \textit{et al.} also reported that hydrogen generation rates depend strongly on temperature, increasing over part of the range from 200 to 300 $^\circ$C before decreasing at higher temperature~\cite{mccollom2016temperature}. These findings indicate that serpentinization is not limited only by equilibrium thermodynamics. Reaction kinetics, mineral texture, transport through secondary phases, and solution chemistry are also important.

The slow kinetics of natural serpentinization motivate active stimulation. Microwave stimulation is attractive because the energy can couple directly to the reacting rock-fluid system, particularly to phases with elevated dielectric loss. In olivine-rich materials, iron-bearing phases and reaction products such as magnetite can alter the effective dielectric response, which may improve coupling as reaction proceeds. This mechanism is distinct from simply raising the bulk temperature of the system and is the central hypothesis of microwave stimulated serpentinization.

\section{Microwave stimulation concept}\label{sec:concept}

\begin{figure}[t]
\centering
\includegraphics[width=0.92\columnwidth]{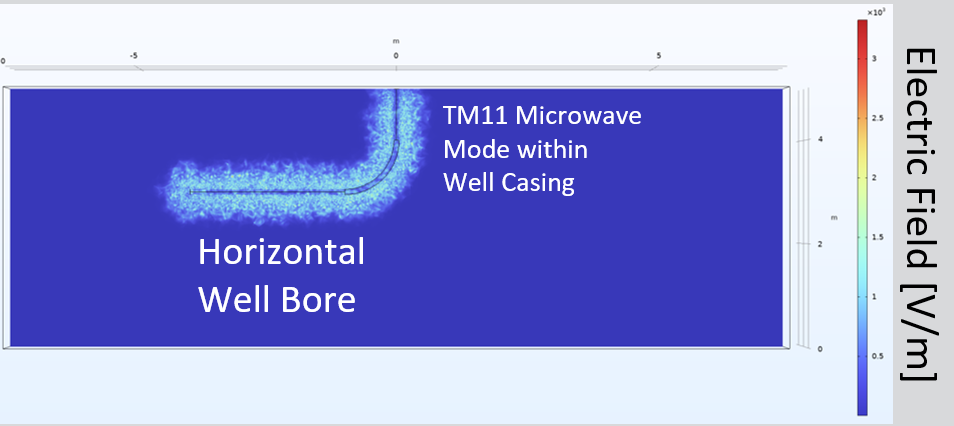}
\caption{COMSOL Multiphysics simulation of a TM11 electromagnetic field injected at the port at top of the domain. The field radiates outwards along the horizontal leg into the dielectric rock. The extent of the field in the rock is limited by the penetration depth.}
\label{fig:efield}
\end{figure}

Figure~\ref{fig:concept} summarizes the broader geologic hydrogen approach that motivates the preliminary experiment. A microwave source is placed at or near the surface, radio-frequency energy is guided through the well bore, and energy is deposited into an olivine-rich formation. The corresponding electromagnetic field is shown in Figure~\ref{fig:efield}, where the simulated field is injected through the port and propagates along the horizontal well-bore geometry into the surrounding dielectric rock. A second well supplies water or brine, and the produced hydrogen-brine mixture is lifted toward the surface using an artificial-lift strategy. In this framework, microwave energy is not necessarily applied continuously. Instead, a pulsed or intermittent stimulus may be sufficient if the exothermic water-rock reaction can be initiated and then sustained by the reaction heat and by continued fluid access.

The broader approach has three linked technical requirements. First, the mineral-fluid system must absorb microwave energy efficiently enough to increase the reaction rate. Second, the enhanced reaction rate must produce hydrogen at a rate that is meaningful relative to the volume of stimulated rock. Third, the net energy balance must be favorable after accounting for microwave generation, transmission losses, pumping or lift energy, and hydrogen recovery. Although those field-scale metrics are outside the scope of the present atmospheric-pressure screening experiment, the laboratory measurement reported here is a necessary first step because any practical deployment requires a reproducible kinetic enhancement relative to conventional heating.

A laboratory article cannot validate the full field-scale energy balance by itself. However, it can provide the critical kinetic evidence needed to determine whether field-scale modeling is justified. The key laboratory outputs are the hydrogen production rate, total hydrogen yield, microwave absorbed power, reflected power, sample temperature history, pressure dependence, solution chemistry, and post-reaction mineralogy. These quantities allow the apparent kinetic enhancement to be compared against the energy input.

\section{Experimental approach}\label{sec:experiments}

\subsection{Materials}

The preliminary feedstock was a prepared ground olivine sample with a nominal 40/70 mesh size. For the initial atmospheric-pressure screening test, approximately 2 g of crushed olivine was saturated with deionized water and placed in a microwave-compatible reaction vessel. Control experiments were performed with otherwise similar sample conditions using conventional hot-plate heating, and additional controls were performed without added water. These controls were intended to distinguish water-rock hydrogen production from sensor response, thermal background, and dry mineral heating.

Future high-pressure experiments should use a defined olivine composition and a controlled particle-size distribution. Because the iron content of olivine affects hydrogen yield, the starting materials should be characterized by X-ray diffraction, scanning electron microscopy, particle-size analysis, and elemental analysis before reaction. A planned test matrix should vary olivine composition, salinity, pH, pressure, and microwave power. A pressure near 500 bar is a useful follow-on target because it approximates relevant subseafloor formation conditions and matches previous serpentinization studies performed at elevated pressure.

\subsection{Microwave reactor and heating method}

The preliminary experiment was conducted 2.45GHz single mode Sairam brand microwave reactor that is composed of a 3kW magnetron source, WR340 waveguides, manual three stub tuner, sliding short and a 1in quartz tub applicator. A follow-on high-pressure program could use an off-the-shelf high-pressure microwave system as a starting platform, followed by development of a custom reaction vessel capable of operation near 500 bar. The pressure vessel design requires mechanical analysis, pressure containment, microwave transparency or coupling control, and safe integration with gas sampling. Finite element analysis should be used to evaluate the pressure-vessel stresses and seals. Electromagnetic simulation should be used to quantify field distribution, reflected power, and power absorbed in the sample and reactor components.

For each run, the microwave power history, reflected power, sample temperature, pressure, gas composition, and experiment duration should be recorded. Because microwave systems can exhibit nonuniform fields and localized heating, the reported microwave input power alone is insufficient. Absorbed power or a calibrated power balance is required to compare microwave heating with conventional heating.

\subsection{Gas measurement and materials characterization}

The preliminary experiment used a metal-oxide gas sensor to measure hydrogen concentration inside or near the reaction vessel. This approach is useful for screening because it provides time-resolved concentration trends, but it is not sufficient as a final quantitative measurement. Follow-on experiments should use mass spectrometry or gas chromatography to quantify hydrogen and other gases. Calibration gases should be used before and after each experimental campaign to account for sensor drift, water-vapor effects, and cross sensitivity.

Post-reaction solids should be characterized by X-ray diffraction to identify serpentine, brucite, magnetite, and residual olivine. Electron microscopy and energy-dispersive spectroscopy should be used to identify reaction rims, secondary phases, and textural changes. Inductively coupled plasma mass spectrometry can be used to quantify dissolved species in the reacted fluid. These measurements are needed to connect hydrogen evolution to mineral conversion rather than only to gas sensor response.

\section{Preliminary results and discussion}\label{sec:results}

\begin{figure}[t]
\centering
\includegraphics[width=0.98\columnwidth]{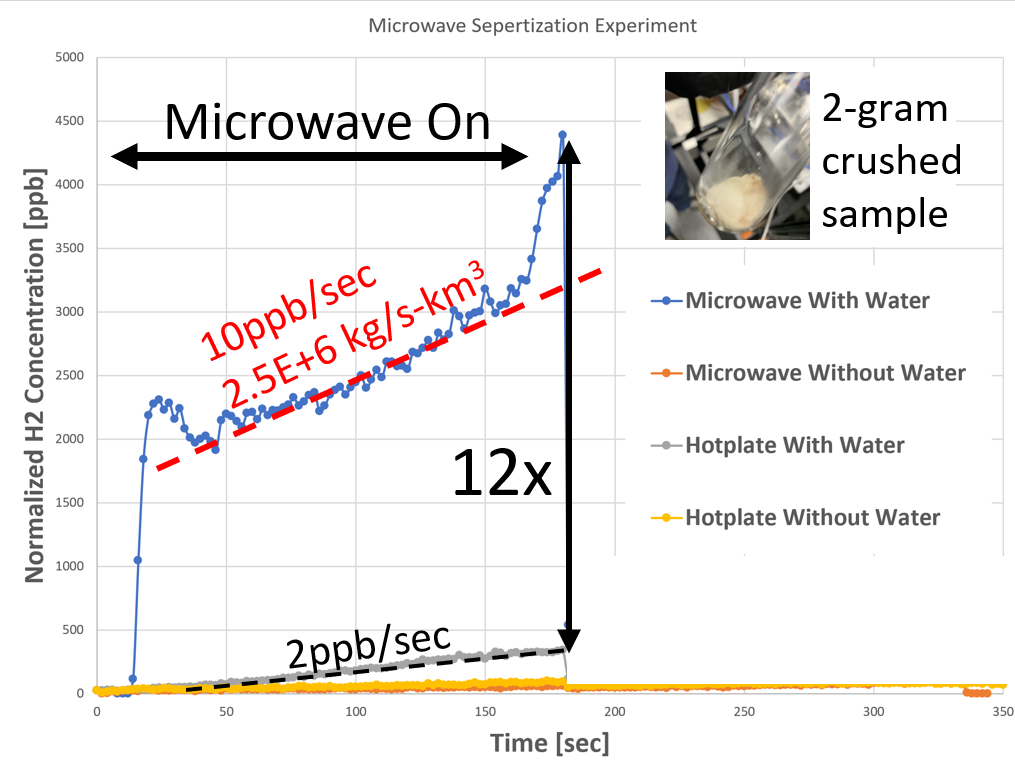}
\caption{Preliminary atmospheric-pressure microwave serpentinization experiment using a 2 g crushed olivine sample with deionized water. The measured hydrogen concentration increased more rapidly under microwave irradiation than under conventional hot-plate heating.}
\label{fig:preliminary}
\end{figure}

Figure~\ref{fig:preliminary} shows the preliminary atmospheric-pressure screening result. The microwave-heated olivine and water sample produced a rapid increase in measured hydrogen concentration after microwave exposure. The 2 g crushed olivine sample showed an approximately 12-fold increase in measured hydrogen concentration relative to the conventionally heated case. Based on the plotted slopes, the microwave and water condition increased at about 10 ppb s$^{-1}$, while the hot-plate and water condition increased at about 2 ppb s$^{-1}$. The dry controls remained comparatively low, indicating that water was necessary for the observed hydrogen signal.

The preliminary result is important because it indicates that microwave exposure changes the hydrogen-generation response beyond the low baseline measured during conventional heating. Several mechanisms could contribute. First, microwave fields can heat the sample volume directly, reducing the thermal lag associated with external heating. Second, selective absorption by iron-bearing phases or reaction products may increase local temperature at reactive interfaces. Third, microwave heating may create microthermal gradients that enhance cracking, fluid access, or transport through altered layers. Fourth, the dielectric response of the sample may evolve as magnetite or other secondary phases form, increasing absorption as reaction proceeds.

However, the preliminary result should be interpreted as a screening observation rather than a complete kinetic measurement. The experiment was performed at 1 atm, while geologic serpentinization is typically considered at elevated pressure. The gas sensor measurement provides concentration rather than absolute moles of hydrogen, so the headspace volume, water vapor, calibration curve, and sensor cross sensitivity affect the quantitative interpretation. In addition, sample temperature and absorbed microwave power must be measured more rigorously to separate a true microwave-specific kinetic effect from a rapid-heating effect. These limitations define the next experimental steps.

\section{High-pressure test matrix and modeling needs}\label{sec:future}

The next stage is to reproduce the preliminary enhancement using calibrated gas measurements and then extend the experiment to formation-relevant pressures. The proposed high-pressure matrix should include microwave and conventional thermal experiments at the same nominal temperature, pressure, water chemistry, and reaction time. The minimum matrix should include the following variables: olivine iron content, particle size, pH, salinity, microwave power, duty cycle, pressure, and temperature. Experiments should also test on/off microwave stimulation to determine whether the reaction can be initiated, paused, or accelerated by controlled electromagnetic exposure.

The experimental program should be coupled to two modeling efforts. The first is a reactor-scale electromagnetic and thermal model. This model should compute electric-field intensity, absorbed power density, sample temperature, and reactor losses. The second is a kinetic and geochemical model. This model should connect hydrogen generation to olivine conversion, iron oxidation, secondary mineral formation, pH, and water activity. The combined model is needed because the measured hydrogen rate is influenced by both chemical kinetics and electromagnetic energy deposition.

A useful reporting metric for later scale-up is hydrogen production rate normalized by reactive rock volume. For the present short communication, the appropriate result is the time-resolved hydrogen concentration response from the 2 g atmospheric-pressure experiment. Conversion to absolute production rate will require calibrated gas analysis, known headspace volume, measured gas flow or pressure change, sample mass, reaction duration, and an estimate of reactive rock density or volume.

\section{Implications for field deployment}\label{sec:deployment}

If microwave stimulation can be shown to increase hydrogen generation at 500 bar while maintaining a favorable energy balance, the approach could complement geothermal and subsurface energy technologies. The well bore can potentially act as a waveguide or delivery path for radio-frequency energy, while an artificial-lift system can assist with moving a hydrogen-bearing brine to the surface. Power recovered from produced hydrogen could offset the microwave energy input, but this claim must be tested through a full energy and techno-economic analysis.

The major field-scale uncertainties are electromagnetic coupling in heterogeneous rock, losses in the well-bore delivery system, fluid transport through low-permeability formations, mineral passivation, hydrogen separation, and long-term control of reaction fronts. These issues are not resolved by the preliminary experiment, but the laboratory result provides motivation for pressure-controlled experiments and model-guided scale-up.

\section{Conclusions}\label{sec:conclusion}

Microwave stimulated serpentinization is a promising but early-stage approach for accelerating geologic hydrogen production from ultramafic rock. The preliminary atmospheric-pressure experiment on water-saturated crushed olivine showed a substantially larger hydrogen sensor response during microwave irradiation than during conventional hot-plate heating. The observed response, including an approximately 12-fold concentration increase and an apparent slope increase from about 2 ppb s$^{-1}$ to about 10 ppb s$^{-1}$, supports the hypothesis that microwave energy can enhance serpentinization kinetics.

The result also identifies the key work needed before the approach can be evaluated for practical deployment. Future experiments must quantify absolute hydrogen yield with calibrated gas analysis, measure absorbed microwave power, characterize mineralogical conversion, and reproduce the effect at elevated pressure near 500 bar. Coupled electromagnetic, thermal, and reaction-kinetics simulations are needed to interpret the experiment and estimate field-scale energy balance. If these measurements confirm that hydrogen production increases faster than microwave energy consumption, microwave stimulation could become a useful pathway for controlled geologic hydrogen production.

\section*{Author declaration}

\subsection*{Conflict of interest}
The authors have no conflicts to disclose.

\subsection*{Author contributions}
A.P. was responsible for formal analysis and writing, review, and editing. T.M. was responsible for conceptualization, formal analysis, supervision, and writing, review, and editing.

\subsection*{Data availability}
The data, programs, and processing scripts that support the findings of this study will be made available upon reasonable request.

\subsection*{Acknowledgment}
The authors acknowledge the West Virginia University Shared Research Facilities and the microwave processing capabilities in the Department of Mechanical, Materials, and Aerospace Engineering.

\bibliography{ref}

@article{kelemen2011rates,
  title={Rates and mechanisms of mineral carbonation in peridotite: natural processes and recipes for enhanced, in situ CO2 capture and storage},
  author={Kelemen, Peter B and Matter, Juerg and Streit, Elisabeth E and Rudge, John F and Curry, William B and Blusztajn, Jerzy},
  journal={Annual Review of Earth and Planetary Sciences},
  volume={39},
  pages={545--576},
  year={2011},
  publisher={Annual Reviews}
}

@article{lamadrid2017effect,
  title={Effect of water activity on rates of serpentinization of olivine},
  author={Lamadrid, Hector M and Rimstidt, J Donald and Schwarzenbach, Esther M and Klein, Frieder and Ulrich, Sarah and Dolocan, Andrei and Bodnar, Robert J},
  journal={Nature Communications},
  volume={8},
  number={1},
  pages={16107},
  year={2017},
  publisher={Nature Publishing Group UK London}
}

@article{schrenk2013serpentinization,
  title={Serpentinization, carbon, and deep life},
  author={Schrenk, Matthew O and Brazelton, William J and Lang, Susan Q},
  journal={Reviews in Mineralogy and Geochemistry},
  volume={75},
  number={1},
  pages={575--606},
  year={2013},
  publisher={Mineralogical Society of America}
}

@article{berndt1996reduction,
  title={Reduction of CO2 during serpentinization of olivine at 300 C and 500 bar},
  author={Berndt, Michael E and Allen, Douglas E and Seyfried Jr, William E},
  journal={Geology},
  volume={24},
  number={4},
  pages={351--354},
  year={1996},
  publisher={Geological Society of America}
}

@article{mccollom2020hydrogen,
  title={Hydrogen generation and iron partitioning during experimental serpentinization of an olivine--pyroxene mixture},
  author={McCollom, Thomas M and Klein, Frieder and Moskowitz, Bruce and Berqu{\'o}, Thelma S and Bach, Wolfgang and Templeton, Alexis S},
  journal={Geochimica et Cosmochimica Acta},
  volume={282},
  pages={55--75},
  year={2020},
  publisher={Elsevier}
}

@article{klein2013compositional,
  title={Compositional controls on hydrogen generation during serpentinization of ultramafic rocks},
  author={Klein, Frieder and Bach, Wolfgang and McCollom, Thomas M},
  journal={Lithos},
  volume={178},
  pages={55--69},
  year={2013},
  publisher={Elsevier}
}

@article{pareek2020insights,
  title={Insights into renewable hydrogen energy: Recent advances and prospects},
  author={Pareek, Alka and Dom, Rekha and Gupta, Jyoti and Chandran, Jyothi and Adepu, Vivek and Borse, Pramod H},
  journal={Materials Science for Energy Technologies},
  volume={3},
  pages={319--327},
  year={2020},
  publisher={Elsevier}
}

@book{international2022global,
  title={Global Hydrogen Review 2022},
  author={International Energy Agency},
  year={2022},
  publisher={OECD Publishing}
}

@article{mazloomi2012electrical,
  title={Electrical efficiency of electrolytic hydrogen production},
  author={Mazloomi, Kaveh and Sulaiman, Nasri B and Moayedi, Hossein},
  journal={International Journal of Electrochemical Science},
  volume={7},
  number={4},
  pages={3314--3326},
  year={2012},
  publisher={Elsevier}
}

@article{osselin2022orange,
  title={Orange hydrogen is the new green},
  author={Osselin, Florian and Soulaine, Cyprien and Fauguerolles, C and Gaucher, EC and Scaillet, Bruno and Pichavant, Michel},
  journal={Nature Geoscience},
  volume={15},
  number={10},
  pages={765--769},
  year={2022},
  publisher={Nature Publishing Group UK London}
}

@article{ajanovic2022economics,
  title={The economics and the environmental benignity of different colors of hydrogen},
  author={Ajanovic, Amela and Sayer, M and Haas, Reinhard},
  journal={International Journal of Hydrogen Energy},
  volume={47},
  number={57},
  pages={24136--24154},
  year={2022},
  publisher={Elsevier}
}

@article{worman2016global,
  title={Global rate and distribution of H2 gas produced by serpentinization within oceanic lithosphere},
  author={Worman, Stacey L and Pratson, Lincoln F and Karson, Jeffrey A and Klein, Emily M},
  journal={Geophysical Research Letters},
  volume={43},
  number={12},
  pages={6435--6443},
  year={2016},
  publisher={Wiley Online Library}
}

@article{mccollom2016temperature,
  title={Temperature trends for reaction rates, hydrogen generation, and partitioning of iron during experimental serpentinization of olivine},
  author={McCollom, Thomas M and Klein, Frieder and Robbins, Mark and Moskowitz, Bruce and Berqu{\'o}, Thelma S and J{\"o}ns, Niels and Bach, Wolfgang and Templeton, Alexis},
  journal={Geochimica et Cosmochimica Acta},
  volume={181},
  pages={175--200},
  year={2016},
  publisher={Elsevier}
}

@article{ford1967high,
  title={High temperature chemical processing via microwave absorption},
  author={Ford, J D and Pei, D C T},
  journal={Journal of Microwave Power},
  volume={2},
  number={2},
  pages={61--64},
  year={1967}
}

@article{agrawal2010latest,
  title={Latest global developments in microwave materials processing},
  author={Agrawal, Dinesh},
  journal={Materials Research Innovations},
  volume={14},
  number={1},
  pages={3--8},
  year={2010},
  publisher={Taylor \& Francis}
}

@article{omran2020microwave,
  title={Microwave catalyzed carbothermic reduction of zinc oxide and zinc ferrite: Effect of microwave energy on the reaction activation energy},
  author={Omran, Mohamed and Fabritius, Timo and Heikkinen, Eetu-Pekka and Vuolio, Tommi and Yu, Yang and Chen, Guo and Kacar, Yilmaz},
  journal={RSC Advances},
  volume={10},
  number={40},
  pages={23959--23968},
  year={2020},
  publisher={Royal Society of Chemistry}
}

@article{huang2012non,
  title={Non-isothermal kinetics of reduction reaction of oxidized pellet under microwave irradiation},
  author={Huang, Z C and Wu, K and Hu, B and Peng, H and Jiang, T},
  journal={Journal of Iron and Steel Research International},
  volume={19},
  number={1},
  pages={1--4},
  year={2012},
  publisher={Elsevier}
}

@article{peng2015microwave,
  title={Microwave-assisted metallurgy},
  author={Peng, Zhiwei and Hwang, Jiann-Yang},
  journal={International Materials Reviews},
  volume={60},
  number={1},
  pages={30--63},
  year={2015},
  publisher={Taylor \& Francis}
}

@article{hassine1995synthesis,
  title={Synthesis of refractory metal carbide powders via microwave carbothermal reduction},
  author={Hassine, N and Binner, J and Cross, T},
  journal={International Journal of Refractory Metals and Hard Materials},
  volume={13},
  number={6},
  pages={353--358},
  year={1995},
  publisher={Elsevier}
}

\end{document}